# What would Plato say? Concepts and notions from Greek philosophy applied to gamification mechanics for a meaningful and ethical gamification

Kostas Karpouzis*1*

*1 Department of Communication, Media and Culture, Panteion University of Social and Political Sciences, Athens, Greece*

**Abstract**
Gamification, the integration of game mechanics in non-game settings, has become increasingly prevalent in various digital platforms; however, its ethical and societal impacts are often overlooked. This paper delves into how Platonic and Aristotelian philosophies can provide a critical framework for understanding and evaluating the ethical dimensions of gamification. Plato's allegory of the cave and theory of forms are used to analyse the perception of reality in gamified environments, questioning their authenticity and the value of virtual achievements, while Aristotle's virtue ethics, with its emphasis on moderation, virtue, and *eudaimonia* (true and full happiness), can help assess how gamification influences user behaviour and ethical decision-making. The paper critically examines various gamification elements, such as the hero's journey, altruistic actions, badge levels, and user autonomy, through these philosophical lenses, and addresses the ethical responsibilities of gamification designers, advocating for a balanced approach that prioritizes user well-being and ethical development over commercial interests. By bridging ancient philosophical insights with modern digital culture, this research contributes to a deeper understanding of the ethical implications of gamification, emphasizing the need for responsible and virtuous design in digital applications.

**Keywords**
Gamification, philosophy, ethics, user experience, autonomy, choice, education 1

## 1. Introduction

In our rapidly evolving digital age, the concept of gamification has garnered significant attention, emerging as a prominent strategy in a wide variety of applications, from education [23] to health [28] and business. Defined by Deterding et al. [8] as the use of game design elements in non-game contexts, gamification seeks to enhance user engagement and motivation through game-like mechanics, often resulting to the modification of user behaviour. The widespread adoption of this strategy has also sparked a considerable amount of academic interest, as evidenced by Hamari, Koivisto, and Sarsa's [14] comprehensive literature review, which scrutinizes the effectiveness and diverse applications of gamification. This paper aims to further this discussion by examining gamification through the lens of ancient Greek philosophy, offering a novel perspective on its implications. Given that different applications of gamification utilise different elements, and motivate or engage users with diverse rewards, we will discuss overarching gamification elements found across different contexts and discuss two specific applications which are successful, yet ethical, as examples.

The writings of ancient Greek philosophers, particularly Plato and Aristotle, provide an insightful framework for exploring many contemporary issues. Arthur [2] underscores the relevance of these philosophers in the 21st century, noting their profound contributions to ethical reasoning and the understanding of human behaviour: Plato's exploration of reality and perception, along with Aristotle's treatises on ethics and human flourishing, offer rich perspectives for analysing modern phenomena. In this work, we discuss the connections between ancient philosophical thought, which influenced numerous subsequent philosophers and thinkers, and modern technological trends, specifically focusing on the implications of gamification in digital applications. Kaplan [19] illustrates the significance of philosophical inquiry in understanding technological phenomena; in the context of gamification, one can explore how the philosophers' insights can illuminate its ethical, psychological, and societal aspects.

For instance, the application of Plato's *theory of forms* and Aristotle's *virtue* ethics to gamification promises to yield novel insights into its ethical and societal impacts [27]. While the growing field of gamification has been extensively explored in terms of user engagement and business models, as Tondello et al. [40] note, there remains a paucity of literature on its philosophical analysis, especially from an ancient Greek standpoint. Revisiting Plato and Aristotle is crucial for understanding gamification in modern society due to their foundational contributions to ethical frameworks, human behaviour understanding, and societal roles (taking also into account the limitations discussed in Section 4.3). Plato's exploration of reality, perception, and ideal forms, alongside Aristotle's detailed virtue ethics and focus on flourishing, provide robust tools for critically evaluating the moral and psychological impacts of gamified systems. Their philosophies encourage a deep examination of whether these digital interactions enhance or detract from genuine well-being, offering timeless principles to navigate the ethical complexities introduced by modern technologies. Thus, their ancient insights are not only historically significant but also immensely relevant in dissecting and guiding the ethical use and design of gamification in contemporary life.

A discussion of ethics in the context of gamification is critically important due to several inherent challenges and potential negative impacts associated with its use [15], including issues related to privacy, the nature of rewards, unfriendly competition, and the exploitation of reward systems. As gamification leverages personal data to customize and enhance user experience, it raises significant privacy concerns; it's essential to ensure that such data is handled ethically, with transparency and respect for user consent [20]. The nature of rewards in gamification can also be ethically ambiguous — while they can motivate and engage, they might also foster addiction or lead users to value virtual achievements over meaningful real-life goals. Furthermore, gamification can sometimes encourage unfriendly competition, promoting a win-at-all-costs attitude that undermines collaboration, community spirit, and fair play. This can be particularly detrimental in educational or work environments where cooperative and supportive interactions are crucial [1]. Lastly, the design of gamified systems often leads to users gaming the system or exploiting loopholes for rewards, which can detract from the actual purpose of the app, whether it be learning, improving health, or another positive goal. These behaviours not only undermine the system's integrity but also raise concerns about encouraging unethical behaviour patterns among users. Therefore, a nuanced discussion of ethics in gamification is vital to guide the design and implementation of these systems, ensuring they are used in ways that respect individual rights and promote overall well-being, rather than detracting from it [38]. This involves a careful balance of engaging users and providing beneficial outcomes while mitigating the risks of misuse, privacy invasion, and negative behavioural incentives.

The structure of this paper is designed to systematically explore the intersection of ancient Greek philosophy and modern gamification. It begins with a theoretical background on key philosophical concepts from Plato and Aristotle, followed by an in-depth analysis of gamification through their perspectives. Subsequent sections discuss the ethical and societal implications of gamification, the responsibility of application designers, and conclude with a critical assessment of the limitations and contemporary relevance of these ancient philosophical viewpoints in the context of modern digital technology.

## 2. Philosophy of gamification

In this section, we go deeper into the writings of Plato and Aristotle, and how those are related to concepts found in different gamification approaches from the points of view of motivation and user autonomy.

### 2.1. Plato's views on reality and perception

Plato's *theory of forms* is a foundational element of his philosophical teachings, where he posits the existence of an abstract world of forms or ideas that represent the truest reality [9]. According to the Greek philosopher, the physical world we experience through our senses is merely a *shadow* of this higher reality. In the realm of gamification, this concept finds a convincing parallel, where the various elements that constitute gamified experiences, such as points, badges, leaderboards, and virtual rewards, can be viewed as the *shadows* or *reflections* of more profound forms of achievement and recognition that exist in the real world. In this context, these gamification elements are mere imitations or approximations of the authentic forms of success, accolade, and accomplishment. The Platonic viewpoint invites a critical examination of the nature and value of these virtual achievements: are they merely superficial imitations of real-world success, or do they hold a more profound significance in the digital age? This perspective challenges us to consider the essence and value of achievements within gamified systems, probing the depth and authenticity of the experiences they offer.

Furthermore, in Plato's *allegory of the cave*, presented in "Republic," [12] he describes a scenario where prisoners, confined in a cave from birth, perceive shadows cast on a wall as the entirety of reality. They remain unaware of the broader world outside the cave, which represents the true reality. This allegory offers another metaphor for the digital realities fashioned by gamified applications, where users immersed in digital environments might come to accept the constructed experiences, narratives, and rewards as their primary reality, much like the prisoners' belief in the shadows as the ultimate truth. This situation brings to the forefront critical questions regarding the nature of reality and illusion: it prompts an exploration of how virtual experiences shape our perceptions and understandings of the world, and whether these digital engagements represent a mere shadow of a more significant, tangible reality. Additionally, it raises the question of *enlightenment* – the journey from the illusion of the virtual world to the broader understanding of reality, akin to the prisoners' eventual ascent from the cave. This Platonic allegory serves as a powerful tool in examining the impact of gamified environments on our perception of reality and our cognitive and psychological engagement with these virtual worlds.

Plato's philosophy placed a strong emphasis on the role of education, envisioning it as a means to attain philosophical wisdom and the pursuit of truth. In "Republic," he proposes an education system geared towards nurturing wisdom and virtuous character. This Platonic vision of education can be intriguingly juxtaposed with contemporary gamified learning applications. These apps utilize game mechanics to potentially enhance educational engagement and learning effectiveness. However, it raises the question of whether such applications align with Plato's ideals of education, which emphasize not just the acquisition of knowledge but the development of critical thinking, philosophical understanding, and the cultivation of virtue. Do these gamified educational tools foster a genuine understanding and wisdom, or do they risk reducing learning to a series of superficial, reward-driven tasks? This consideration is vital in evaluating the effectiveness and philosophical alignment of gamification in education, probing whether such tools truly contribute to the formation of well-rounded, critically thinking individuals as envisioned by Plato.

In examining the ethical dimensions of gamification through a Platonic lens, one must also consider Plato's profound concern with virtue and the health of the soul. Plato's ethical philosophy [21], with its emphasis on the pursuit of goodness and the development of moral virtues, offers a significant framework for evaluating the ethical implications of gamified environments. These digital platforms, particularly those that involve elements of social interaction, competition, and reward mechanisms, present various ethical challenges [11]. Issues such as fairness, the impact on individual character, and the promotion or undermining of virtuous behaviour become crucial considerations [39]. How do these gamified systems align with Plato's vision of ethical living and the cultivation of a harmonious soul? Do they promote ethical behaviour and contribute positively to the individual's moral development, or do they risk encouraging less virtuous traits, such as competitiveness at the expense of fairness or collaboration? This exploration is essential in understanding the ethical impact of gamification, assessing whether these digital experiences contribute to or detract from the development of a virtuous character as advocated by Plato.

Plato's influence extends beyond his immediate philosophical contributions, shaping subsequent perspectives on various aspects of life, including play and games. Plato addresses the notion of playing connected to the notion of knowledge in his dialogue "Theaetetus" [31], but also other philosophical ideas of his have significantly influenced later philosophical and cultural views on these topics [37]. In the context of gamification, it is insightful to consider how later philosophers, drawing on Platonic thought, have approached the concept of play and games. For instance, the balance between reality and illusion, a key theme in Plato's works, is central to understanding the nature of digital games and gamified experiences. Furthermore, the ethical considerations surrounding competitive play and the role of games in education can be traced back to Platonic ideals about the pursuit of truth, the development of virtue, and the nature of reality. This historical and philosophical perspective enriches the understanding of gamification, providing a deeper insight into its roots in classical thought and its implications in the modern world.

## 2.2. Aristotle's ethics and virtue theory

Aristotle, Plato's student, contributed extensively to various applied and theoretical sciences, including ethics, politics, metaphysics, and logic. His ethical framework was centred around the concept of *virtue ethics* and is a cornerstone of his philosophical contributions. The overarching Aristotelian notion of *eudaimonia*, i.e., true and full happiness, can be connected to three gamification mechanics: the reward system, engagement, and the end goal of gamification itself: in his "Nicomachean Ethics," [6] Aristotle posits that virtues are habits or qualities that lead to a good life (which he defines as a life of *eudaimonia*). Virtues, for Aristotle, are the mean between two extremes – excess and deficiency [21]. When considering gamification, this framework prompts an examination of how gamified elements influence the development or erosion of virtues. For instance, do the reward systems in gamified apps encourage moderation and balance, or do they lead to excessive behaviour such as addiction? The application of virtue ethics to gamification raises critical questions about the moral development of users engaged in these digital environments and whether these platforms contribute to or hinder their pursuit of a virtuous, flourishing life.

The Aristotelian concept of eudaimonia, often translated as 'happiness' or 'flourishing,' is a pivotal aspect of his ethical theory. Aristotle argues that eudaimonia is achieved through a life of virtue and rational activity, fulfilling one's potential. In the context of gamification, this raises intriguing questions about the role of digital apps in promoting or impeding eudaimonia. Does engagement with gamified apps lead to a fulfilling and meaningful life as envisioned by Aristotle, or does it detract from it? For instance, educational apps that employ gamification could potentially align with Aristotle's vision by enhancing learning and personal growth. However, if the gamification in these apps focuses merely on superficial rewards or addictive mechanics, it may deviate from the Aristotelian ideal of fostering genuine well-being and personal development.

Aristotle's *teleology*, the study of purpose or end goals, is another crucial aspect of his philosophy. He believed that everything has a purpose or end to which it naturally aims [35]. This perspective can be applied to gamification by examining the purposes and end goals of gamified systems. Are these systems designed with the end goal of enhancing user well-being, or are they primarily oriented towards commercial gain or user addiction? Aristotle's teleological framework encourages an analysis of the ultimate objectives behind gamification strategies, questioning whether they align with the promotion of the good life and fulfilment of human potential.

In his ethical treatises, Aristotle also delves into the nature and role of pleasure in human life. He distinguishes between higher and lower forms of pleasure [33], suggesting that true happiness is found in intellectual and virtuous activities rather than in the mere satisfaction of desires. This distinction is particularly relevant in examining gamified apps, which often use pleasure as a primary motivator for user engagement. The challenge is to discern whether the pleasures offered by these apps are aligned with Aristotle's higher pleasures, contributing to the user's intellectual and moral development, or whether they cater to lower, more transient forms of pleasure, potentially leading to addictive behaviours and a diversion from more meaningful pursuits.

## 2.3. Social aspects: the role of community

Aristotle's philosophy places significant emphasis on the role of community and social relationships in the pursuit of eudaimonia. In his view, articulated notably in his work "Politics," [25] humans are by nature social beings, and a fulfilling life can only be achieved within the context of a community. Aristotle believed that ethical virtues are largely developed through and within social interactions and communal relationships. This concept is particularly relevant when examining social gamification platforms and multiplayer gaming environments.

In these digital contexts, community interaction can take various forms, from collaborative tasks and team-based challenges to competitive play and social networking. The Aristotelian lens prompts us to inquire whether these platforms foster a sense of genuine community, mutual respect, and collective well-being, in line with Aristotle's vision of communal life contributing to personal and collective flourishing [30]. For example, when gamified apps encourage teamwork, collaborative problem-solving, and positive social interactions, they align with the Aristotelian ideal by enhancing social bonds and contributing to the common good. These platforms can become spaces where players not only engage in entertainment but also develop virtues like cooperation, empathy, and generosity, contributing to their moral growth.

Conversely, if gamification leads to negative social behaviours such as isolation, excessive competition, or cyberbullying, it would be antithetical to Aristotle's view of the community's role in ethical development. It's crucial to analyse how the design and mechanics of these gamified environments impact the nature of social interactions. Are they fostering healthy, supportive communities, or are they fostering toxic competitive environments? This perspective urges us to consider how gamified environments can be structured to enhance social welfare and promote a sense of community that contributes to the overall well-being of its members.

Moreover, Aristotle's concept of the *polis* (city-state) as a framework for ethical development underscores the importance of designing gamified environments that not only entertain but also enrich the social and moral fabric of the community. In this context, developers and designers of gamified systems have a responsibility to consider the social and ethical implications of their designs, ensuring that these platforms contribute positively to the communal and individual well-being.

## 2.4. Practical wisdom and decision-making

Aristotle's concept of *phronesis*, a lived, practical wisdom that involves understanding and reflecting on the right course of action in varied situations, is a crucial aspect of his virtue ethics, emphasizing the ability to make good judgments and decisions that lead to a flourishing life. In "Nicomachean Ethics," Aristotle describes phronesis not just as theoretical knowledge, but as a lived, practical wisdom that involves understanding the right course of action in varied situations [17]. When we apply this concept to gamified environments, it challenges us to examine whether these platforms encourage the development of such practical wisdom. Do gamified applications and systems encourage users to make thoughtful, well-considered decisions, or do they foster impulsive, reward-driven behaviours? This analysis is essential in understanding how gamification impacts users' decision-making processes, and whether it aligns with the Aristotelian ideal of cultivating a life of virtue and practical wisdom [4].

The design of gamified environments often incorporates elements that influence decision-making processes, such as immediate rewards, feedback loops, and competitive structures. These elements can potentially skew users' decision-making towards *short-term gains* and *instant gratification*, which may conflict with Aristotle's concept of practical wisdom that emphasizes *long-term well-being* and *moral virtue*. The challenge lies in assessing whether gamified systems encourage users to engage in reflective thinking, consider the consequences of their actions, and make decisions that contribute to their overall well-being and personal growth [41]. For instance, in educational gamification, the focus should ideally be on encouraging learners to make decisions that foster a deeper understanding and mastery of the subject matter, rather than merely pursuing points or badges.

A critical aspect of practical wisdom in the context of gamification is balancing the allure of immediate rewards with the pursuit of long-term goals and virtues [36]. Aristotle's ethics suggest that true happiness and flourishing come from activities that fulfil our rational nature and contribute to personal and moral growth. This raises the question of whether gamified systems, with their emphasis on immediate rewards and achievements, potentially detract from this longer-term perspective. The design of these systems, therefore, holds significant ethical implications: it should ideally foster an environment where users are encouraged to pursue activities that are not only rewarding in the short term but also conducive to their long-term personal, moral and cultural development, in the perspective of a lifelong learning, as proposed by UNESCO Institute for Lifelong Learning (UIL) [42].

The application of Aristotle's concept of practical wisdom extends to the social and collaborative aspects of gamification. In multiplayer or social gamified platforms, users often face decisions that affect not only their individual experience but also the community as a whole [26]. This scenario offers a unique opportunity to examine how gamification can promote or hinder the development of social virtues and wise decision-making in a communal context. Do these platforms encourage cooperation, fairness, and consideration of others' well-being, or do they promote selfish, short-sighted behaviours? Aristotle's perspective would advocate for gamified environments that nurture a sense of community, ethical interaction, and decisions that reflect not only personal good but also the common good.

## 3. Implications to gamification elements

Exploring individual gamification elements and their relation to philosophical concepts requires an understanding of both the mechanics of gamification and the philosophical principles they may embody or contradict. In this Section, we delve into several gamification elements, extending beyond traditional elements, such as points, badges, and leaderboards, to consider more complex and meaningful aspects like the hero's journey, altruistic actions, badge levels, and autonomy.

The fundamental aspects of gamification include game mechanics (points, badges, leaderboards), dynamics (progress, relationships, emotions), and aesthetics (visual design, story, player experience). These elements create an engaging and often immersive experience for users. From a Platonic perspective, these mechanics can be seen as shadows or imitations of real-world experiences, while from an Aristotelian view, they can be analysed for their impact on virtues and the pursuit of eudaimonia. In addition, gaming can be paralleled with concepts such as Plato's forms and Aristotle's notions of virtue and purpose. The competitive and strategic aspects of gaming reflect Aristotle's emphasis on virtues like courage and wisdom, achieved through balanced actions and decisions. Similarly, the immersive worlds of digital games can be viewed through Plato's lens of reality and perception, questioning the nature of reality in digital experiences versus the physical world. These parallels form a rich tapestry for understanding the philosophical underpinnings of gaming and gamification.

Moving to more advanced gamification concepts, the hero's journey, a narrative framework often employed in gamified environments, resonates deeply with Platonic philosophy. In Plato's works, particularly in his dialogues, there is a recurrent theme of an individual's journey towards knowledge and enlightenment, akin to the hero's journey in mythological narratives. This journey often involves overcoming challenges, gaining wisdom, and ultimately achieving a form of self-actualization [5]. In gamified contexts, this can translate to users embarking on a metaphorical journey through the app or platform, facing challenges, improving their skills with respect to app usage, and achieving goals or results that lead to personal growth. The hero's journey in gamification mirrors Plato's allegory of the cave, where the journey from illusion to enlightenment is a central theme. This narrative structure can elevate the gamified experience from mere entertainment to a more meaningful process of personal development, aligning with Platonic ideals of knowledge and self-realization.

Furthermore, incorporating elements of altruism and helping other users in gamified systems aligns closely with Aristotle's ethics, particularly his emphasis on the development of moral virtues through social interaction. In "Nicomachean Ethics," Aristotle posits that virtues are developed through practice and are oriented towards the good of others as well as oneself. Gamification mechanisms that encourage users to help others, whether through collaborative tasks, shared challenges, or supporting fellow users, cultivate virtues like generosity, empathy, and kindness. This not only enhances the individual user's experience but also contributes to a communal sense of well-being and interconnectedness, resonating with Aristotle's concept of eudaimonia as a communal, as well as personal, achievement.

Finally, the Pursuit of Excellence, as implemented through badge levels in gamification, can reflect the Aristotelian pursuit of excellence or *arete*, provided it has been designed thoughtfully [35]. In Aristotle's view, excellence is achieved through the continual practice and refinement of virtues. Similarly, badge levels in gamified systems can represent stages of personal and skill development, encouraging users to strive for higher levels of competence and understanding. This aspect of gamification can be particularly effective in educational contexts, where badge levels can signify not just achievement but also a deeper mastery of knowledge and skills. This aligns with the Aristotelian view that true happiness comes from fulfilling one's potential and excelling in one's endeavours.

## 4. Ethical implications

From a Platonic perspective, the ethical concerns in gamification arise when the virtual achievements and rewards (like points and badges) create an illusion of success and fulfilment. Plato, in his allegory of the cave, warned against mistaking shadows for reality. In the context of gamification, this allegory can be interpreted as a caution against valuing virtual achievements over genuine accomplishments. When users are driven to extreme behaviours like overspending or manipulating ('gaming') the system for more rewards, they are, in a Platonic sense, mistaking the 'shadows' of success (virtual rewards) for real achievement; this misalignment can lead to addictive behaviours and a distorted sense of value, which is ethically concerning as it moves the individual away from the pursuit of true knowledge and genuine fulfilment.

Aristotle's virtue ethics, centred on finding the mean between excess and deficiency, provides a framework for evaluating the ethical implications of gamification. When users engage in extreme spending or exploit systems for more points, they are exhibiting behaviours that Aristotle would classify as excesses, which are contrary to the development of moral virtues. Such actions reflect a lack of temperance, a key virtue in Aristotle's ethical system. The purpose of gamification, from an Aristotelian perspective, should be to promote balance and virtuous behaviour, contributing to the user's overall well-being (eudaimonia). When gamified systems incentivize excessive or unethical behaviours, they are failing to align with this ethical standard, potentially leading users away from the path of virtue and well-being.

## 4.1. The role of designers in ethical gamification

The ethical responsibility in gamification also extends beyond the users, to the designers and creators of these systems. Both Plato and Aristotle emphasized the role of educators and leaders in shaping the moral and intellectual character of individuals. In the modern context, this translates to a responsibility on the part of gamification designers to create systems that encourage ethical behaviour and personal growth. Designers should be mindful of the potential for their systems to encourage addictive or unethical behaviours and strive to create balanced and ethical gamified experiences. This involves designing reward systems that promote healthy engagement and discourage harmful extremes, aligning with the Platonic ideal of pursuing true understanding and the Aristotelian goal of fostering virtue and well-being.

While gamification is often employed as a marketing tool to increase user engagement and product loyalty, it is crucial to consider the ethical implications of such strategies. Plato and Aristotle would both argue that the primary goal of any educational or engagement tool should be the betterment and improvement of the individual. When gamification is used primarily for marketing purposes, with little regard for the user's well-being or ethical development, it can lead to exploitative practices, such as encouraging excessive spending or competitive behaviours that do not align with the development of virtue or the pursuit of genuine happiness [41]. Ethical gamification should prioritise the user's improvement and well-being above marketing objectives, ensuring that engagement strategies are aligned with fostering positive, virtuous behaviours and experiences.

To mitigate the potential for harmful behaviours in gamified systems, a balanced approach that incorporates both Platonic and Aristotelian ethical principles is essential [26]. This involves creating gamification strategies that encourage users to *reflect* on their actions (a Platonic ideal), promoting an understanding of the difference between virtual achievements and real-world accomplishments. Simultaneously, these systems should be designed to foster Aristotelian virtues, such as *moderation* and ethical decision-making, discouraging behaviours that lead to *excesses*. This ethical framework necessitates a thoughtful design process where user well-being is a primary consideration, ensuring that gamification serves as a tool for positive personal development rather than as a means to exploit user vulnerabilities for commercial gain [3].

## 4.2. Good practices

Duolingo is a renowned language-learning application which has successfully incorporated gamification to increase and retain user interest and engagement. Central to Duolingo's success is its adaptive learning algorithm, which employs machine learning to personalize the learning journey for each individual user. By tracking users' progress and adjusting the difficulty of lessons and exercises based on their performance, this algorithm ensures a tailored and optimized learning experience. In addition, the ability to choose their own learning path, offers users the sense of autonomy and self-sufficiency; this is discussed by Aristotle in his *Nicomachean Ethics* as the ideal for humanity, as being self-sufficient makes our lives choiceworthy and lacking in nothing.

The foundation of Duolingo's gamification strategy lies in a triumvirate of core elements: Points, Levels, and Rewards; these game-like features serve as motivational tools to keep learners consistently engaged and inspired without rewarding excessive use or allowing learners to 'game the system' by engaging with activities merely for the reward or to accumulate points. Duolingo also uses the concepts of "streaks" to motivate users to stay committed to their language learning goals: a streak is the number of consecutive days a user has achieved their daily goal. Studies have shown [10] that repeated studies of the same material over several days influences its subsequent retention. Regarding more advanced gameful elements, recent versions of Duolingo include a "Stories" feature, which introduces interactive dialogues that simulate real-life conversations; this allows learners to navigate through these dialogues, applying their acquired language skills in context. Duolingo also fosters interaction among its users through social features, enabling users to exchange tips and advice and even engage in language practice with native speakers.

Finally, Duolingo is known for its constant notifications, which serve as reminders or encouragements for users to continue their language learning. These notifications aim to make the user feel valued and wanted on the platform, igniting a sense of 'calling' to the user; this is strongly related to the concept of motivation involving reason and desire, as noted by Aristotle in *Nicomachean Ethics* and *De Anima* [16].

Another practice of meaningful gamification is described by Legaki et al. in [24] and [22]. The authors developed a game called "Horses for Courses", as part of an experiment in the context of statistics education. The game incorporates main game design patterns related to challenge-based gamification, including Points, Levels, Challenges, as well as a leaderboard, fostering motivation and illustrating progress in the framework of the course. Relating this to the philosophies of Aristotle and Plato, we can draw some parallels, as both philosophers emphasised the importance of challenges and learning. Aristotle, in his *Nicomachean Ethics*, discussed the concept of *virtue*, which involves finding the balance between deficiencies and excesses. This can be seen as a parallel to the challenges in the game, which require players to find a balance between different strategies to succeed.

In addition, Plato, in his work *'The Republic"*, proposed the idea of a just society where individuals perform roles that they are naturally suited to. This can be seen as a parallel to the game's leaderboard, which recognises and rewards players based on their performance and skills.

Finally, in the context of autonomy, "Horses for Courses" can be seen as a recognition of individual autonomy and the ability of individuals to make choices that best suit their skills and preferences. This aligns with Aristotle's concept of self-sufficiency (autarkeia). However, it's important to note that while there are parallels, the concepts of "Horses for Courses" and the philosophies of Aristotle and Plato are from vastly different contexts and times. The application of ancient philosophical concepts to modern games should be done thoughtfully and with an understanding of these differences.

### 4.3. Limitations

While the philosophies of Plato and Aristotle provide insights into human nature, ethics, and society, their application to modern contexts is not without challenges. The limitations of their works stem from the historical and cultural specificity of their times, the evolution of societal norms and values, and the advent of technologies that redefine human experience [34]. Adapting their philosophies to contemporary societies requires a critical and creative engagement with their ideas, an openness to reinterpretation, and a thoughtful consideration of the diversity and complexity of modern life. As societies continue to evolve, so too must our understanding and application of these ancient philosophical insights [7]. For instance, while the allegory of the cave provides a powerful metaphor for understanding enlightenment and the pursuit of truth, its applicability to modern issues such as media literacy, digital echo chambers, and virtual realities necessitates a re-evaluation, especially when it comes to the concept of "the good" which seems to be forgotten by Plato [32]. The simplistic dichotomy between the enlightened and unenlightened may not fully capture the complexities of modern information consumption and the varied experiences of digital engagement. In addition, Aristotle's virtue ethics is based on the cultivation of moral virtues through moderation and rationality. However, modern societies are characterized by a diversity of cultural values, ethical systems, and life experiences. The universal applicability of Aristotelian virtues may be limited, and adapting his ethical framework to contemporary multicultural and pluralistic societies requires an inclusive approach that accommodates diverse conceptions of the good life. Finally, Aristotle's teleological view of nature and purpose assumes inherent goals and purposes in all beings and systems. In the age of Artificial Intelligence and complex technological and societal systems, the notion of inherent purposes becomes problematic. Adapting this view requires a rethinking of purpose and intentionality in systems that are human-made and often driven by multiple, sometimes conflicting, objectives.

## 5. Conclusions

In this paper we examined the philosophical implications of gamification in digital applications through the perspectives of ancient Greek philosophers Plato and Aristotle. Through a detailed analysis, it was revealed how Plato's theory of forms and the allegory of the cave can be instrumental in understanding the nature of reality as presented in gamified environments, highlighting concerns about the illusion of success and the potential disconnection from authentic experiences. Aristotle's virtue ethics, with its focus on moderation, virtue, and the pursuit of eudaimonia, provided a framework for assessing the impact of gamification on user behaviour, particularly concerning the development or erosion of virtues like temperance and ethical decision-making.

The exploration of various gamification elements such as the hero's journey, altruistic actions, badge levels, and autonomy, within these philosophical contexts, underscored the potential for both positive personal growth and the risk of negative behaviours like addiction and ethical manipulation. The ethical responsibilities of designers were also emphasized, suggesting that gamification systems should prioritize user well-being and moral development over commercial gains.

Moving forward, several areas warrant further exploration:

- Cross-Cultural Philosophical Perspectives: Future research could include the analysis of gamification through other philosophical traditions, such as Eastern philosophies or contemporary approaches to autonomy and identity (e.g., by Sartre or Nietzsche), to provide a more diverse and comprehensive understanding of its implications.
- Empirical Studies: There is a need for empirical research to observe and measure the actual impact of gamified systems on user behaviour, particularly in relation to the development of virtues and ethical decision-making as proposed by Aristotle.
- Technological Advancements and Gamification: As technology evolves, particularly with the advent of augmented reality (AR) and virtual reality (VR), the ethical implications of more immersive gamified experiences should be explored. These developments offer new realms for applying and testing ancient philosophical insights in modern contexts.
- Ethical Guidelines for Gamification Design: Developing comprehensive ethical guidelines for gamification designers, rooted in philosophical principles, could help in creating more responsible and beneficial gamified systems.
- Long-Term Impact Studies: Longitudinal studies examining the long-term impacts of gamification on individuals' psychological well-being, societal behaviour, and perception of reality would provide valuable insights into the sustained effects of these digital engagements.

In conclusion, while gamification presents numerous opportunities for enhancing user engagement and experience, it is imperative that its design and implementation are guided by ethical considerations that prioritize human well-being. The insights from Plato and Aristotle provide a valuable framework for understanding and navigating the ethical landscape of gamification, but they also open the door for further philosophical explorations and empirical research in this ever-evolving field.

# Acknowledgements

This work has been funded by the Horizon Europe Road-STEAMER project (Grant number: 101058405). More information can be found at https://www.road-steamer.eu